# Accurate determination of the valence band edge in hard x-ray photoemission spectra using GW theory

Johannes Lischner
jlischner597@gmail.com
Materials Sciences Division, Lawrence Berkeley National Laboratory, Berkeley 94720, USA.
Department of Physics and Department of Materials and the Thomas Young Centre for Theory and Simulation of Materials, Imperial College London, London SW7 2AZ, United Kingdom.

Slavomír Nemšák
Department of Physics, University of California, Davis, California 95616, USA, and Materials Sciences Division, Lawrence Berkeley National Laboratory, Berkeley 94720, USA.
Peter-Gruenberg-Institut-6, Forschungszentrum Juelich, Juelich, Germany.

Giuseppina Conti
Department of Physics, University of California, Davis, California 95616, USA, and Materials Sciences Division, Lawrence Berkeley National Laboratory, Berkeley 94720, USA.

Andrei Gloskovskii
Deutsches Elektronen-Synchrotron DESY, Photon Science, Notkestraße 85, D-22607 Hamburg, Germany.

Gunnar Karl Pálsson
Department of Physics, University of California, Davis, California 95616, USA, and Materials Sciences Division, Lawrence Berkeley National Laboratory, Berkeley 94720, USA.
Present address: Department of Physics and Astronomy, Uppsala University, Uppsala, SE-751 20 Sweden.

Claus M. Schneider
Peter-Gruenberg-Institut-6, Forschungszentrum Juelich, Juelich, Germany.

Wolfgang Drube
Deutsches Elektronen-Synchrotron DESY, Photon Science, Notkestraße 85, D-22607 Hamburg, Germany

Steven G. Louie
Department of Physics, University of California, Berkeley, California 94720, USA, and Materials Sciences Division, Lawrence Berkeley National Laboratory, Berkeley 94720, USA.

Charles Fadley
Department of Physics, University of California, Davis, California 95616, USA, and Materials Sciences Division, Lawrence Berkeley National Laboratory, Berkeley 94720, USA.

**Abstract:**

We introduce a new method for determining accurate values of the valence-band maximum in x-ray photoemission spectra. Specifically, we align the sharpest peak in the valence-band region of the experimental spectrum with the corresponding feature of a theoretical valence-band density of states curve from ab initio GW theory calculations. This method is particularly useful for soft and hard x-ray photoemission studies of materials with a mixture of valence-band characters, where strong matrix element effects can render standard methods for extracting the valence-band maximum unreliable. We apply our method to hydrogen-terminated boron-doped diamond, which is a promising substrate material for novel solar cell devices. By carrying out photoemission experiments with variable light polarizations, we verify the accuracy of our analysis and the general validity of the method.

*Introduction.*---An accurate knowledge of the valence-band offset $\Delta E_{VBO}$ of a semiconductor heterostructure is needed to understand transport properties across the interface and electron-hole recombination rates, which determine the efficiency of many semiconductor devices, such as thin-film solar cells. Following Kraut and coworkers[1,2], the valence-band offset at an interface between two semiconductors (labeled A and B) can be expressed as

$$\Delta E_{VBO} = E_{VBM}^A(i) - E_{VBM}^B(i) = E_{CL}^A(i) - E_{CL}^B(i) - [(E_{CL}^A - E_{VBM}^A) - (E_{CL}^B - E_{VBM}^B)], \qquad (1)$$

where $E_{CL}^A$ and $E_{VBM}^A$ denote the binding energy of a shallow core level and the valence-band maximum in the bulk of semiconductor A, respectively, and $E_{VBM}^A(i)$ denotes the binding energy of the valence-band maximum at the interface. $E_{CL}^B$, $E_{VBM}^B$ and $E_{VBM}^B(i)$ represent the same quantities for semiconductor B. Finally, $E_{CL}^A(i) - E_{CL}^B(i)$ denotes the energy difference of the shallow core levels of semiconductors A and B at the heterojunction. Equation (1) assumes that each core level rigidly follows the valence-band maximum of the material as the junction is formed, and it has been used in numerous studies, including of metal-oxide heterostructures[3].

Photoemission spectroscopy probes the energies of occupied electronic states of a material and can provide all the quantities needed to evaluate Equation (1). However, interpreting experimental photoemission spectra can be challenging because of the inherent complexities of the photoemission process and the difficulty in extracting accurate values of $E_{VBM}$[3]. The theoretical description of photoemission experiments is an area of active research and simple pictures, such as the three-step model, where the photoemission process is described as an absorption step, a transport step and a surface transmission step, often fail to reproduce experimental observations[4,5].

Interpretations of experimental photoemission spectra are thus often guided by theoretical band structure calculations. For example, Kraut and coworkers[1,2] proposed fitting a theoretical model, which assumes that the photoemission spectrum is proportional to a broadened valence-band density of states (VB DOS) obtained from band structure calculations, to the measured spectrum in the vicinity of the valence-band edge and determining $E_{VBM}$ from the fit. Typically, density-functional theory (DFT) within a Kohn-Sham formulation is used to calculate band structures and the VB DOS[3,6,7]. While such calculations can give important insights into the experimental spectra, their accuracy depends on the approximation for the exchange-correlation energy. Highly accurate band structures can be obtained from the ab initio GW method[8,9]. In this approach, the electron self energy, which captures many-electron interaction effects beyond mean-field theory, is expressed as the product of the interacting Green's function G and the screened Coulomb interaction W.

As a newer development in experiment, hard x-ray photoemission spectroscopy (HXPS) employs photon energies in the multi-keV range[10,11]. This greatly reduces the surface sensitivity (from which photoemission spectroscopy at lower photon energies suffers) and makes the study of interfaces buried 100 Å or more below the surface possible. Because of these advantages, HXPS is becoming an increasingly popular technique for the experimental determination of valence-band offsets[12,13]. However, at such high photon energies, matrix element effects can play a crucial role and can cause significant errors in valence-band offsets determined using standard methods, as we will show below.

In this paper, we introduce a simple new method for combining experiment with GW theory for determining valence-band maxima using HXPS, or more broadly, conventional XPS in the keV regime. Instead of fitting a theoretical model to the experimental spectrum in the vicinity of the valence-band edge, we align the sharpest peak in the valence-band region of the experimental spectrum with the corresponding feature in a theoretical VB DOS curve of the quasiparticles from a GW theory calculation. We apply our method to hydrogen-terminated boron-doped diamond (HBD), which is a promising substrate material in novel solar cell devices[13–15]. We further verify the accuracy of the new method by carrying out measurements with variable light polarizations[16].

*Methods.*--- The standard method to determine the binding energy of the valence-band edge from experimental XPS spectra of semiconductors is due to Kraut and coworkers[1,2]. They suggested that the measured spectrum in the vicinity of the valence-band edge is proportional to the VB DOS, obtained from band structure calculations and broadened to account for the experimental energy resolution. Fitting the theoretical model

$$I(E) = S\, N_v(E - \Delta E) + B, \qquad (2)$$

to the experimental spectrum finally yields the energy of the valence-band edge. Here, $N_v(E)$ denotes the broadened VB DOS (whose energies are referenced to the theoretical $E_{VBM}$), $S$ is a scale factor, $B$ accounts for a constant random-noise background and $\Delta E$ denotes an energy shift. Treating $S$, $B$ and $\Delta E$ as fitting parameters, accurate values for $E_{VBM}$ could be estimated for semiconductors[1,2].

The applicability of this method rests on two assumptions: (i) the photoemission spectrum is proportional to the broadened VB DOS and (ii) the chosen method of band structure theory produces an accurate VB DOS. For many materials, assumption (i) is valid in the vicinity of the valence-band edge for soft x-ray photons, but becomes problematic in the HXPS regime, where strong matrix element effects can suppress or enhance the contributions of specific electronic states to the experimental spectrum.

When photoemission occurs from states with different matrix elements, the resulting spectrum can differ significantly from the VB DOS. In this case, one might attempt to determine the valence band edge by fitting a matrix element weighted VB DOS given by

$$I(E) = \sum_{i\alpha} S_{i\alpha} N_{i\alpha}(E - \Delta E) + B .\qquad(3)$$

to the measured spectrum. Here $N_{i\alpha}(E)$ denotes the contribution to the VB DOS from the atomic state $i$ of the atomic species $\alpha$ and the scale factors $S_{\alpha i}$ are in first approximation proportional to the differential *atomic* photoelectric cross section $d\sigma_{i\alpha}/d\Omega$. This approximation to the photoemission spectrum becomes increasingly more accurate at higher photon energy[17], but fitting this model to an experimental spectrum is more complicated because of the larger number of parameters.

Regarding assumption (ii), we note that the accuracy of VB DOS curves from DFT calculations[3,18] depends on the approximation for the exchange-correlation energy functional. In contrast, the ab initio GW method[8,9,19] yields accurate quasiparticle energies and consequently VB DOS curves for many weakly and moderately correlated semiconductors and insulators.

Based on the accuracy of the ab initio GW method, we propose the following method to determine $E_{VBM}$: Instead of fitting a theoretical model to the experimental photoemission spectrum *in the vicinity of $E_{VBM}$*, we locate the sharpest peak in the valence-band region of the experimental spectrum and align it with the corresponding feature in the theoretical VB DOS curve from ab initio GW theory calculations. This fixes the energy scale of the theoretical model and immediately yields an estimate for $E_{VBM}$. The resulting method is less sensitive to matrix element effects, which can strongly influence the spectrum near the valence band edge, and does not suffer from high noise levels arising from low photoemission signals in the onset region.

However, the sharpest peak of the spectrum is not necessarily located in the vicinity of the valence-band edge and consequently a band structure method is required, which accurately captures the energies of *all occupied valence-band quasiparticle states*. To achieve this, we employ the GW method. For diamond, Jimenez and coworkers demonstrated that the GW method gives good agreement with the experimental bandwidth, while DFT within the local density approximation results in a significant underestimation[20]. We also note that the proposed method contains only a *single alignment parameter*, the position of the sharpest peak in the experimental spectrum.

In what follows, we also compare our results for $E_{VBM}$ with results obtained using the method of Chambers et al.[3], who observed that the approach of Kraut and coworkers results in significant inaccuracies for complex oxide materials. Chambers et al. proposed to fit both the onset of the experimental spectrum near the valence band edge and the background noise (dark counts) to straight lines and then determine $E_{VBM}$ from the intersection of these two lines. While this method is very simple to use and does not require the calculation of a theoretical VB DOS, its validity depends on the assumption of a linear VB DOS in the vicinity of $E_{VBM}$.

*Experimental and computational details*.--- We apply our method to synthetic hydrogen-terminated boron-doped diamond (HBD) from Element Six Ltd. with boron concentration ≈$10^{20}$ cm$^{-3}$ (570 ppm). We determine the energy separation of the valence-band maximum and the carbon 1s (C 1s) core state, which is needed to calculate the valence-band offset of a heterostructure consisting of HBD and another semiconductor. We measure HXPS spectra of HBD using photon energies of 2.5 keV and 5.9 keV, thus varying the effective sampling depth in the experiment.

A first set of HXPS measurements were performed at 2.5 keV and 5.0 keV at the bend-magnet beam line 9.3.1 of the Advanced Light Source, Berkeley, USA. The end station was equipped with a Scienta SES-2002 hemispherical analyzer upgraded to operate at up to 6 keV, and using a multichannel plate-phosphor-CCD detection scheme. The total energy resolution of these measurements was 0.4 and 0.7 eV for the photon energies 2.5 and 5.0 keV, respectively. These measurements were carried out with p-polarized light. To investigate the role of matrix element effects, we also carried out HXPS measurements with variable light polarizations using 5.9 keV photons. For this, we used the undulator beam line P09 of PETRA III, DESY, Hamburg, Germany[21]. Two linear polarizations (s- and p-) were employed using a diamond phase retarder[22] to selectively enhance or suppress contributions from different electronic states to the photoemission spectrum. The spectra in these measurements were collected using a Specs Phoibos 225 analyzer with a delay-line detector. The total resolution was here kept at 0.2 eV. All photon energies were calibrated using gold as a reference. In order to maximize the photoelectron yield, all experiments were conducted in grazing incidence/normal emission geometries. Thus, with p-polarization, the electric field is very near to the axis of the entry lens into the spectrometer and maximizes the intensity from s-character. By contrast, with p-polarization, the field is very nearly perpendicular to the emission direction of the photoelectrons, strongly suppressing intensity from s-character. However, there is also a strong dependence on photon energy that tends to suppress C 2p relative to C 2s at higher energies, and both effects must be considered[23]. We illustrate this below with plots of C 2s and C 2p differential photoelectric cross sections.

(Note that s- and p- in polarization, are distinctly different from the s- and p- in orbital character.)

To obtain theoretical VB Kohn-Sham DOS curves, we carried out DFT calculations with the Perdew, Burke, Ernzerhof (PBE) exchange-correlation energy as implemented in the QUANTUM ESPRESSO program package[24]. We employed a norm-conserving pseudopotential and a plane wave basis with an 80 Ry energy cutoff. We then computed the quasiparticle energies using the GW method as implemented in the BerkeleyGW program package[25]. The static dielectric matrix was computed on a 16x16x16 k-point grid in the Brillouin zone using 80 empty states and an 8 Ry dielectric cutoff. We also employed the modified static remainder correction of Deslippe and coworkers to include additional empty states in the self-energy[26]. The generalized plasmon-pole model of Hybertsen and Louie[9] was used to extend the static dielectric matrix to finite frequencies. We obtain a total occupied valence bandwidth of 23.06 eV, in excellent agreement with the experimental value of $23.0 \pm 0.2$ eV[20]. The occupied Kohn-Sham bandwidth in the DFT calculation (without GW theory corrections) is 21.50 eV, significantly smaller than the experimental value.

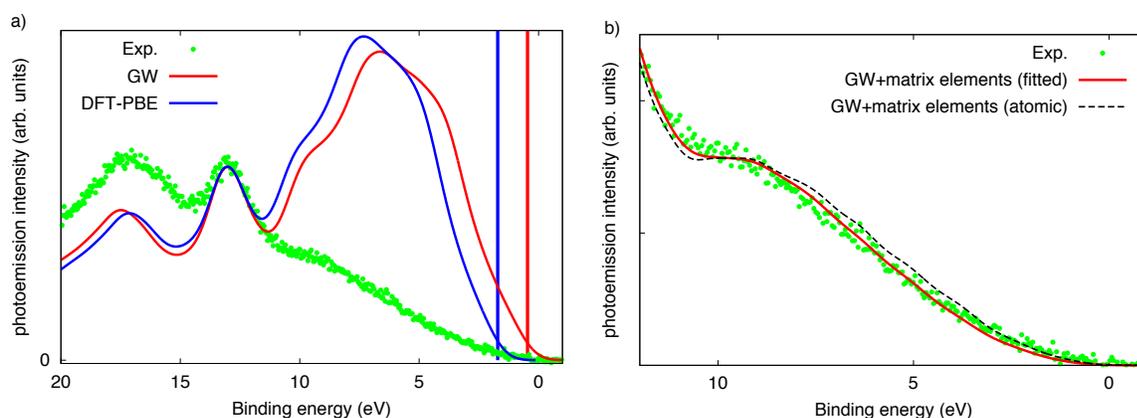

**Figure 1(a): X-ray photoemission spectrum of hydrogen-terminated boron-doped diamond obtained using p-polarized 5.0 keV photons (green dots) and theoretical valence-band density of states curves from DFT (blue line) and GW (red line). The theoretical curves were convolved with a Gaussian of 0.7 eV width to account for the experimental broadening. The vertical lines mark the positions of the valence band maximum estimated from DFT (blue) and GW theory (red). (b): Comparison of the experimental photoemission spectrum (green dots) and the theoretical spectrum including fitted matrix-element effects from GW theory (red line) in the vicinity of the valence band edge. Also, shown is the result when atomic cross sections are used to allow for matrix-element effects (black dotted line).**

*Results.---* Figure 1(a) shows the experimental photoemission spectrum obtained using p-polarized photons with an energy of 5.0 keV. We observe two prominent peaks, one sharp peak at a binding energy of 12.56 eV and a broader peak at 17.1 eV. Also shown are theoretical VB DOS curves from DFT-PBE and GW theory, which were convolved with a Gaussian function with a full width at half maximum of 0.7 eV (determined by fitting the gold 4f line of a reference sample) to account for the experimental broadening. Both theoretical curves exhibit a large, broad peak at low binding energies [between 10 and 0 eV, see Fig. 1(a)]. This peak arises mostly from states with a strong contribution from carbon 2p (C 2p) orbitals.

Figure 2 shows the partial atomic photoemission cross sections $\frac{d\sigma_{nl}}{d\Omega} = \frac{\sigma_{nl}}{4\pi}[1 + \beta_{nl} P_2(cos\theta)]$ of C 2p and C 2s orbitals for a range of photon energies. Here, $\beta_{nl}$ denotes an energy-dependent asymmetry parameter, $\theta$ is the angle between the electric field polarization and the photoelectron wave vector (which is parallel to the surface normal in our setup) and $P_2(x) = (3x^2 - 1)/2$. The cross section $\sigma_{nl}$ of C 2p orbitals decreases much more rapidly with increasing photon energy than the one for C 2s orbitals[23]. For p-polarized photons with an energy of 5.0 keV, this results in an enhancement of photoemission from C 2s orbitals by more than a factor of 100 compared to C 2p orbitals. It is important to note that the asymmetry parameter for C 2s states is very close to 2 and varies only weakly with photon energy. For s-polarized photons with $P_2(0) = -1/2$, this leads to a strong suppression of photoemission from C 2s orbitals, as noted above.

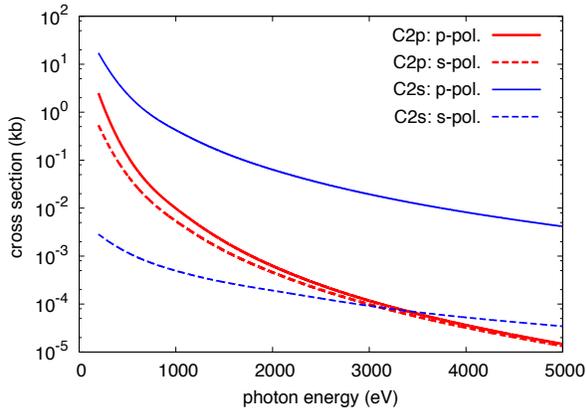

**Figure 2: Atomic photoemission cross sections of C 2p (red) and C 2s (blue) orbitals for p-polarized (solid lines) and s-polarized (dashed lines) light.**

Because of its sharpness, the peak at 12.56 eV binding energy constitutes an ideal reference point to align the experimental spectrum and the theoretical VB DOS curves. Specifically, we have shifted the VB DOS curve until the position of the maximum of the sharp peak agrees with the maximum of the corresponding feature in the experimental spectrum. Figure 1(a) also shows the resulting estimates for the position of the VBM as vertical lines. The estimates from GW and DFT differ by 1.25 eV, a result of the significant underestimation of the bandwidth in DFT. From the experimental spectrum, we also extract the energy of the HBD C 1s core level, $E_{CL}$, and find its separation from the valence-band edge to be

$E_{CL} - E_{VBM} = 283.74$ eV (4).

We repeated the measurement and analysis with a photon energy of 2.5 keV and obtained $E_{CL} - E_{VBM} = 283.71$ eV, in excellent agreement with the value obtained at the higher photon energy. This result demonstrates that our method to extract $E_{VBM}$ is robust and does not suffer from inaccuracies when matrix element effects are important. The extracted values for $E_{CL} - E_{VBM}$ agree quite well with the result of Shi and coworkers[27], who obtained $E_{CL} - E_{VBM} = 283.58$ eV using the method of Chambers et al., but disagrees by 0.6 eV from the value obtained by Liu and coworkers[28], $E_{CL} - E_{VBM} = 283.10$ eV. These authors also employed the method of Chambers et al. to extract $E_{VBM}$.

For the energy of the valence-band edge relative to the Fermi level, we find a value of 0.45 eV for 5.0 keV photons and 0.21 eV for 2.5 keV photons. We attribute the difference to the sum of surface band bending effects[29] resulting from an increased escape depth of photoelectrons and an increased recoil effect arising from the larger photon momentum[30].

Having obtained an accurate value for $E_{VBM}$, we determine the full spectrum by computing the matrix element weighted VB DOS [Equation (3)]. For this, we project the DFT wave functions of the diamond crystal onto atomic orbitals and thus obtain the separate contributions to the total VB DOS from C 2s and C 2p orbitals, which we denote $N_{C2s}$ and $N_{C2p}$, respectively. We then determine the scale factors $S_{C2s}$ and $S_{C2p}$, which describe matrix element effects, through a least-squares fit to the experimental spectrum. Figure 1(b) shows the resulting theoretical spectrum, which agrees well with experiment. Also shown is a theoretical spectrum for which we have simply used the theoretical atomic cross sections in Eq. (3). Again, we find good agreement with experiment.

*Comparison to other methods for extracting $E_{VBM}$.*--- To understand why previous studies reported significantly different values for the separation of the C 1s core level from the valence-band maximum, we also extract $E_{VBM}$ from the experimental photoemission spectrum using the methods of Kraut et al.[1,2] and Chambers et al.[3].

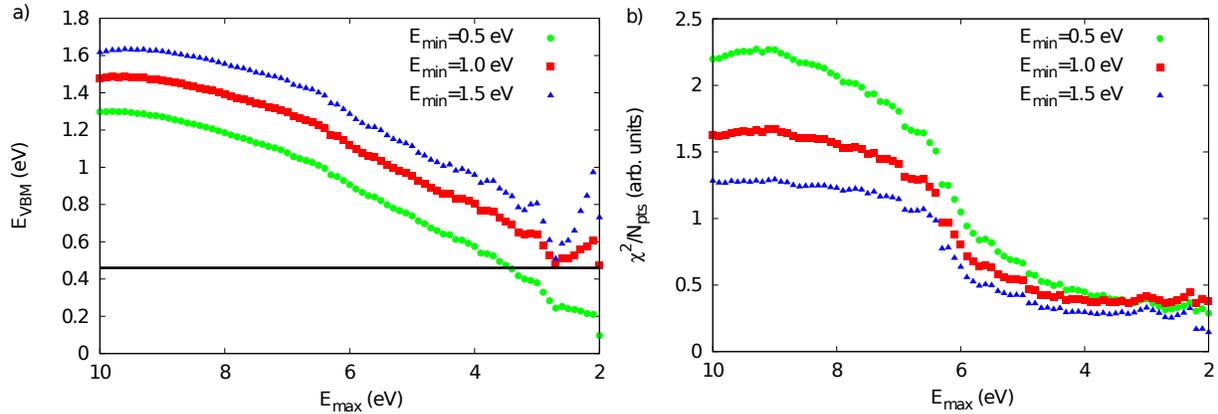

**Figure 3(a): Estimates of the valence-band maximum of diamond obtained from the method of Chambers and coworkers. Different binding energy domains $[E_{min}, E_{max}]$ were used for the linear fit to the experimental spectrum. The horizontal black line denotes the value of $E_{VBM}$ obtained by aligning the sharpest peak of the experimental photoemission spectrum with the corresponding feature of the valence-band density of states from GW theory. (b): Corresponding values of the figure-of-merit $\chi^2$ divided by the number of data points contained in the fitting domain.**

We first attempted to use the method of Chambers and coworkers[3] to determine the valence-band maximum. After subtracting a constant noise background, a straight line is fitted to the experimental spectrum in a binding energy domain between $E_{min}$ and $E_{max}$ in the vicinity of the valence-band edge and $E_{VBM}$ is determined by the intersection of the straight line with the binding energy axis. Figure 3(a) shows that resulting estimates of $E_{VBM}$ depend sensitively on the binding energy domain used for the fit. Also, we do not find a clear minimum of the figure-of-merit $\chi^2$ [see Figure 3(b)]. It is thus not clear how to extract a meaningful value of $E_{VBM}$ using the method of Chambers and coworkers.

Next, we used the method of Kraut et al.[1,2] [see Equation (2)]. This approach is based on the assumption that the photoemission spectrum is proportional to the broadened VB DOS in the vicinity of the valence band edge. To test the validity of this assumption, we computed the ratio of $N_{C2p}$ multiplied by $S_{C2p}$, the atomic photoemission cross section of the C 2p orbital, and $N_{C2s}$ multiplied by $S_{C2s}$ (see Figure 4). At a photon energy of 1.5 keV, there is clearly a region near the valence band maximum where 2p states dominate. In this region, the photoemission spectrum is approximately proportional to the VB DOS. At 5.0 keV, however, the cross section ratio of s- and p-states is $\frac{S_{C2s}}{S_{C2p}} >$ 100 [see Figure (2)] and even very close to the VBM C 2s states give an important contribution to the photoemission spectrum. Consequently, the spectrum is *not proportional* to the VB DOS and the standard method of Kraut cannot be applied to diamond in the HXPS regime because of the important role of matrix element effects. We also attempted to use a modified version of this method and fitted a matrix element weighted VB DOS to the experimental spectrum. However, we again find that the results depend sensitively on the fitting domain (see appendix for details).

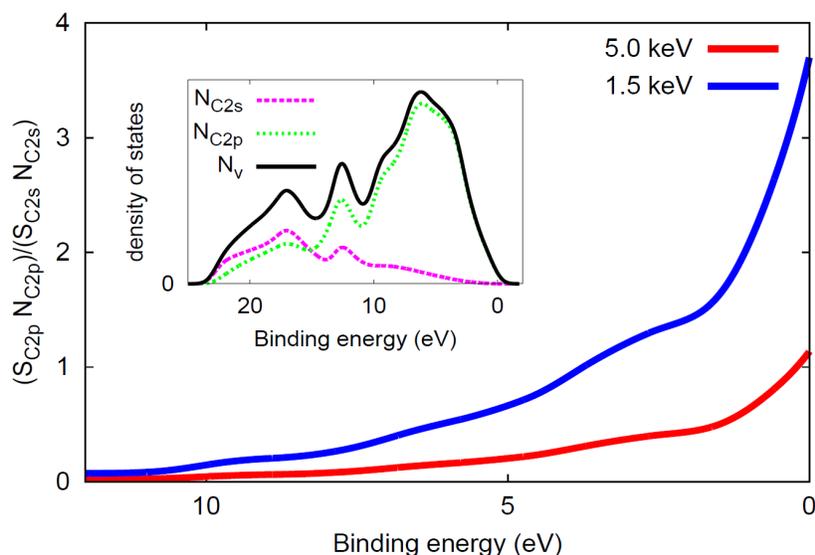

**Figure 4: Ratio of matrix-element weighted carbon 2s and carbon 2p projected density-of-states curves for diamond at photon energies of 1.5 keV (blue) and 5.0 keV (red). The inset shows the total valence-band density of states (black) of diamond from GW theory and the contributions from carbon 2s (magenta) and 2p (green) states.**

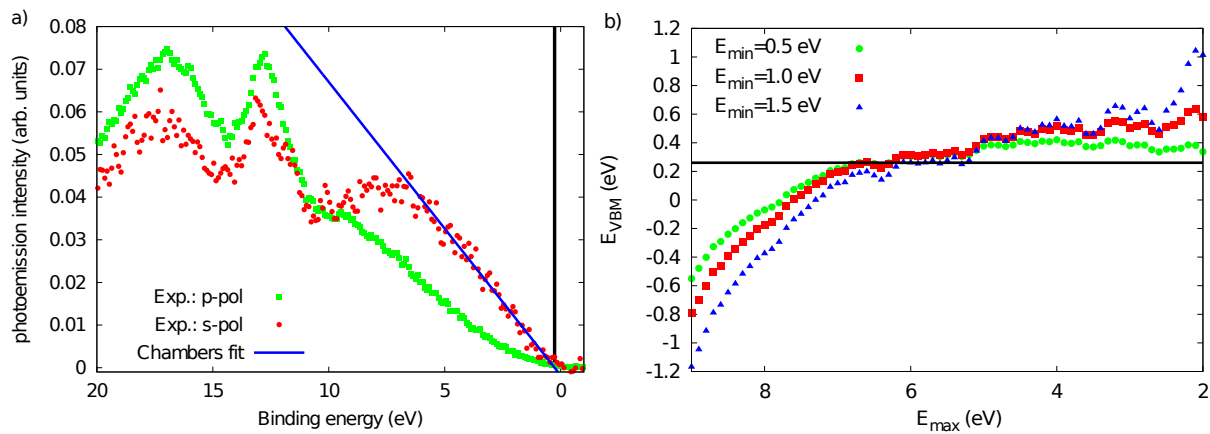

**Figure 5(a): Experimental photoemission spectra of diamond using s-polarized (red dots) and p-polarized (green dots) 5.9 keV photons. Also shown is the linear fit to the spectrum obtained using a binding energy fitting domain of [5,1] eV. The resulting estimate for the valence band maximum agrees well with the value obtained by aligning the sharpest peaks in the GW valence band density of states with the corresponding features in the experimental spectrum (black vertical line). (b): Estimates of the valence band maximum from the method of Chambers and coworkers for different binding energy fitting domains $[E_{max}, E_{min}]$. The black horizontal line denotes the value of $E_{VBM}$ obtained from GW theory.**

*Comparison to photoemission with variable light polarization.*--- To verify the accuracy of our new method for determining $E_{VBM}$, we carried out HXPS measurements with variable light polarizations using 5.9 keV photons. Figure 2 shows that the use of s-polarized light dramatically reduces the cross section of C 2s states, while the cross section for C 2p states is reduced only by a small amount. Therefore, for s-polarized light the photoemission spectrum of HBD is more similar to the broadened VB DOS and standard methods for extracting $E_{VBM}$ should be applicable. We demonstrate now that this is indeed the case.

Figure 5(a) shows the experimental photoemission spectra obtained from s- and p-polarized light. For s-polarized photons, the relative contribution of states with p-character is greatly increased in the experimental spectrum. We first apply the method of Chambers et al.[3] to the experimental spectrum obtained using s-polarized photons. Figure 5(b) shows that the resulting estimates of $E_{VBM}$ still depend on the binding energy domain used for the fit, but much less than for p-polarized photons [Figure 4(b)]. In particular, all curves in Figure 5(b) are relatively flat and nearly coincide for $E_{max}$~6 eV, indicating a robust fit. The value of $E_{VBM}$ in this region also agrees well with the value obtained by aligning sharp peaks in the VB DOS from GW theory and the experimental spectrum [black horizontal line in Figure 5(b)]. Similarly, we observe that the method of Kraut and coworkers yields a more robust fit and the resulting value for $E_{VBM}$ is in good agreement with the other methods (see appendix).

*Conclusions.*--- We have introduced a new method for the accurate determination of the valence-band maximum in experimental x-ray photoemission spectra. By aligning the sharpest peak in the experimental spectrum with the corresponding feature in the quasiparticle valence-band density of states curve from GW theory, accurate values of the valence-band maximum can be extracted. This method is particularly well suited for the hard x-ray regime, where strong cross section/matrix element effects can cause significant inaccuracies of standard methods for determining the valence-band maximum, even though using variable x-ray polarization can to some degree compensate for these effects. We applied the new method to hydrogen-terminated boron-doped diamond and determined an accurate value of the energy separation of the carbon 1s core state and the valence-band maximum, which is needed for the calculation of heterojunction band discontinuities and Schottky barrier heights. Finally, we verify the accuracy of our approach by carrying our measurements with variable light polarizations.

*Acknowledgments.*--- This work was supported by the SciDAC Program on Excited State Phenomena (methods and software developments) and Theory Program (GW calculations) funded by the U. S. Department of Energy, the Office of Basic Energy Sciences and of Advanced Scientific Computing Research, under Contract No. DE-AC02-05CH11231 at the Lawrence Berkeley National Laboratory, and by NSF Grant No. DMR15-1508412 (theoretical analysis). Computational resources have been provided by the DOE at NERSC. GKP, and CSF acknowledge salary support from the U.S. Department of Energy, Office of Science, Office of Basic Energy Sciences, Division of Materials Sciences and Engineering under Contract No. DE-AC02-05CH11231, via the LBNL Materials Sciences Division, Magnetic Materials Program, and Grant DE-SC0014697, via the Dept. of Physics, University of California Davis. GKP also acknowledges the International Union for Vacuum Science, Technique and Applications and the Swedish Research Council for financial support as well as partial salary support and travel support from a MURI grant of the Army Research Office (Grant No. W911-NF-09-1-0398). Funding for the photoemission instrument at beamline P09 (DESY) by the German Federal Ministry of Education and Research (BMBF) under contracts 05KS7UM1 and 05K10UMA with Universität Mainz; 05KS7WW3, 05K10WW1 and 05K13WW1 with Universität Würzburg is gratefully acknowledged.

Appendix: Alternative fitting procedure

*Modified method of Kraut and coworkers.---* As shown in the main text, at high photon energies the photoemission spectrum of diamond in the vicinity of the valence-band edge is no longer proportional to the VB DOS because of matrix element effects. This renders the original methods of Kraut and coworkers for extracting $E_{VBM}$ inapplicable. Instead, we attempted to extract $E_{VBM}$ by fitting a matrix element weighted VB DOS [Equation (3)] to the experimental spectrum in Figure 1(a).

Specifically, for each value of $\Delta E$ in Equation (3), we determine the scale factors $S_{\alpha i}$ by a least-squares fit to the experimental spectrum. The accuracy of the fit can be judged by the figure-of-merit $\chi^2$ defined as

$$\chi^2(\Delta E) = \sum_{i_{min}}^{i_{max}} \{I_{exp}(E_i) - I_{fit}(E_i, \Delta E)\}^2, \qquad (A1)$$

where $I_{exp}$ denotes the experimental spectrum, $E_i$ the measured binding energies and $I_{fit}(E_i, \Delta E)$ the theoretical spectrum for a given value of $\Delta E$ obtained by optimizing all other parameters in Equation (3). The indices $i_{min}$ and $i_{max}$ establish a domain of binding energies used for the least-squares fit. Figure A1 shows $\chi^2(\Delta E)$ curves for three different fitting domains. We observe that the minimum of the $\chi^2$ curve and thus the estimated value of $E_{VBM}$ depends sensitively on the fitting domain rendering any conclusions based on this method subject to high errors of $\pm 0.4$ eV.

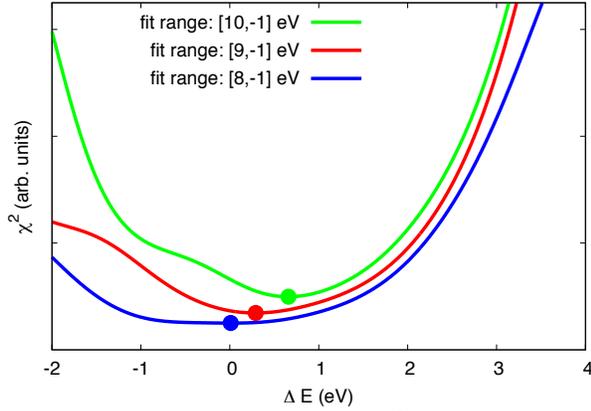

**Figure A1:** Figure-of-merit $\chi^2$ for the fit of Equation (3) to the experimental photoemission spectrum at 5.0 keV photon energy. Different binding energy domains were used for the fitting: $E_B$=[10,-1] eV (green), $E_B$=[9,-1] eV (red) and $E_B$=[8,-1] eV (blue). The dots denote the estimated valence band maxima.

*Application to photoemission with variable light polarization.---* We applied the method of Kraut and coworkers [Equation (2)] to the experimental spectrum obtained using s-polarized photons [Figure 5(a)]. Figure A2(a) shows that the fit is now more robust and less dependent on the binding energy fitting domain than for p-polarized photons [Figure A1]. Figure A2(b) shows that the experimental spectrum agrees well with a shifted VB DOS from GW theory calculations and the resulting estimate of $E_{VBM}$ is very close to the value obtained by aligning the sharpest peaks in the GW VB DOS and the experimental spectrum.

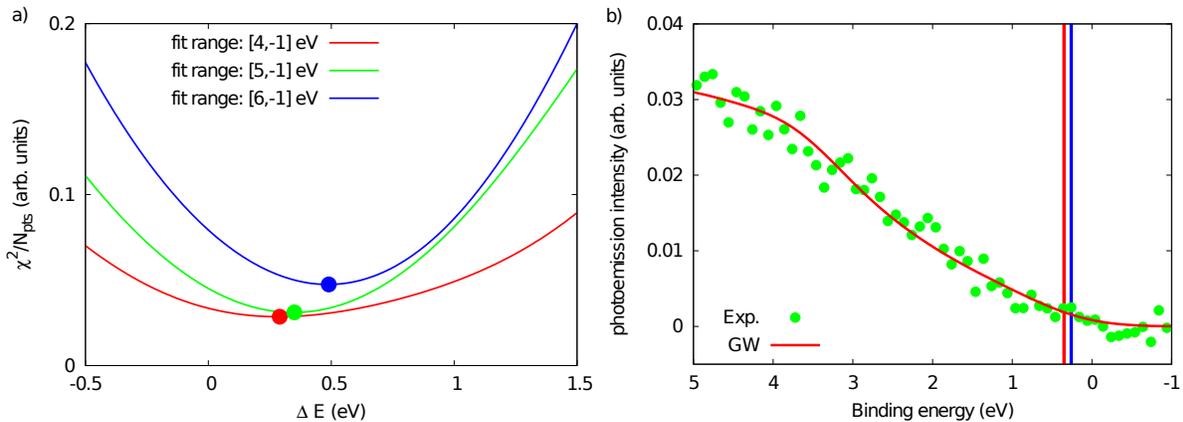

**Figure A2(a):** Figure-of-merit $\chi^2$ for fitting the valence band density of states from GW theory to the experimental photoemission spectrum using s-polarized 5.9 keV photons in the vicinity of the valence band edge (method of Kraut and coworkers). (b): Comparison of the resulting shifted valence band density of states from GW theory for a binding energy fitting domain [5,-1] eV and the experimental photoemission spectrum. The vertical red line denotes the resulting estimate for the valence band maximum and the blue vertical lines denotes the estimate obtained by aligning the sharpest peak in the

**GW valence band density of states with the corresponding feature in the experimental spectrum.**